\documentclass[aps,showpacs,superscriptaddress,nofootinbib,floatfix]{revtex4-1}
\usepackage{graphicx}
\usepackage{epsf}
\def\slashchar#1{\setbox0=\hbox{$#1$}
   \dimen0=\wd0 \setbox1=\hbox{/} \dimen1=\wd1
   \ifdim\dimen0>\dimen1 \rlap{\hbox to \dimen0{\hfil/\hfil}} #1
   \else  \rlap{\hbox to \dimen1{\hfil$#1$\hfil}} / \fi}
\newcommand{\beas}{\begin{eqnarray*}}
\newcommand{\eeas}{\end{eqnarray*}}
\newcommand{\bea}{\begin{eqnarray}}
\newcommand{\eea}{\end{eqnarray}}

\begin{document}

\title{$\bar B_s\to K$ semileptonic decay from an Omn\`es improved constituent 
quark model} 

\author{C.  \surname{Albertus}} 
\affiliation{Departamento
de F\'\i sica At\'omica, Nuclear y Molecular\\  e Instituto Carlos
 I de F\'{\i}sica Te\'orica y Computacional\\ Universidad de Granada, Avenida
 de Fuentenueva s/n, E-18071
Granada, Spain.} 
\author{E.  \surname{Hern\'andez}} 
\affiliation{Departamento
de F\'\i sica Fundamental e IUFFyM,\\ Universidad de Salamanca, Plaza de la
Merced s/n, E-37008
Salamanca, Spain.}  
\author{C.  \surname{Hidalgo-Duque}} 
\affiliation{Instituto de F\'\i sica Corpuscular (IFIC), Centro Mixto
  CSIC-Universidad de Valencia, Institutos de Investigaci\'on de
  Paterna, Apartado 22085, E-46071 Valencia, Spain} 
\author{J.  \surname{Nieves}}
\affiliation{Instituto de F\'\i sica Corpuscular (IFIC), Centro Mixto
  CSIC-Universidad de Valencia, Institutos de Investigaci\'on de
  Paterna, Apartado 22085, E-46071 Valencia, Spain} 
  

\today

\begin{abstract}
We study the $f^+$ form factor for the semileptonic $\bar B_s\to K^+\ell^-\bar\nu_\ell$
decay in a constituent quark model.  The valence quark estimate is
supplemented with the contribution from the $\bar B^*$ pole that dominates
the high $q^2$ region. We use a multiply-subtracted
Omn\`es dispersion relation to extend the quark model predictions from its
region of applicability near $q^2_{\rm max}=(M_{B_s}-M_K)^2\sim 23.75$ GeV$^2$ to all $q^2$
values accessible in the physical decay. To better constrain the dependence of
$f^+$ on $q^2$, we fit the subtraction constants to a combined input
from previous light cone sum rule [Phys.\ Rev.\ D {\bf 78} (2008)
  054015] and the present quark model results.  From this analysis, we obtain
$\Gamma(\bar B_s\to K^+\ell^-\bar\nu_\ell)=(5.45^{+0.83}_{-0.80})|V_{ub}|^2\times
10^{-9}\,{\rm MeV}$, which is about 20\% higher than the prediction
based only on QCD light cone sum rule estimates. Differences are much
larger for the $f^+$ form factor in the region above $q^2=15$ GeV$^2$.
\end{abstract}

\maketitle

\section{Introduction}

The magnitude of the $V_ {ub}$ element  of the
Cabibbo-Kobayashi-Maskawa (CKM) quark mixing matrix plays a critical
role in testing the consistency of the Standard Model (SM) of particle
physics and, in particular, the description of CP violation. Any
inconsistency could be a sign of new physics beyond the SM. $V_{ub}$
is currently the least well-known element of the CKM matrix and
improvement in the precision of its determination is highly desirable
and topical.  At present, there exist some tension between the
$|V_{ub}|$ values extracted from the analysis of inclusive decays and
those from the study of exclusive channels. Thus, for instance, BABAR
measurements of the inclusive electron and photon spectra in the $B\to
X_ue\nu_e$ and $B\to X_s\gamma$ decays were used in
Ref.~\cite{Golubev:2007cs} to extract $|V_{ub}|$ from data. Two
different methods were used that led to two different values for the
magnitude of the CKM matrix element $V_{ub}$,
$(4.28\pm0.29\pm0.29\pm0.26\pm0.28)\times10^{-3}$ and
$(4.40\pm0.30\pm0.41\pm0.23)\times10^{-3}$, respectively.  These
estimates could be compared with the value of
$(3.41^{+0.37}_{-0.32}|_{\rm th} \pm 0.06|_{\rm exp}) \times10^{-3}$
obtained in Ref.~\cite{Khodjamirian:2011ub} from the exclusive
semileptonic $B\to\pi$ form factors computed in QCD light cone sum
rules (LCSR) for $q^2< 12$ GeV$^2$ and the latest BABAR data available
at the time.

In general, the determinations based on inclusive semileptonic decays
using different calculational ans\"atze are consistent. The largest
parametric uncertainty comes from the error on the $b-$quark mass. The
PDG 2013 update~\cite{Beringer:1900zz} (review by R. Kowalewski and
T. Mannel) quotes an inclusive average $|V_{ub}| = (4.41 \pm
0.15^{+0.15}_{-0.17}) \times10^{-3}$.  The value obtained from
exclusive determinations, largely dominated by the semileptonic
$B\to\pi$ decay, and quoted in ~\cite{Beringer:1900zz} is $|V_{ub}| =
(3.23 \pm 0.31)\times10^{-3}$, where the precision is limited by form
factor normalizations\footnote{The value obtained from the semileptonic
  exclusive $B\to\rho$ decay is even smaller by around  20\%
  \cite{Flynn:2008zr,delAmoSanchez:2010af}. However, it has been
  recently pointed out \cite{Meissner:2013pba} that $\pi\pi$
  distribution effects in the broad $\rho-$width, not considered
  in \cite{Flynn:2008zr,delAmoSanchez:2010af}, might
  lead to an enhanced $|V_{ub}|$ value determined from this
  decay mode.}. The two determinations are independent, but are marginally
consistent with each other (see also the  discussion and averages
provided by the heavy flavor averaging group (HFAG) in
\cite{Amhis:2012bh}\footnote{A complete listing of the averages and
  plots, including updates since \cite{Amhis:2012bh} was prepared, are
  also available on the HFAG web site~\cite{HFAG}.}).  On the other
hand, $|V_{ub}|$ is also determined by the UTfit
collaboration~\cite{UTfit} from the unitarity triangle analysis within
the SM. The tension between exclusive and inclusive
determinations of the $|V_{ub}|$ CKM matrix element is now playing a
major role in these SM fits because of the increased accuracy on
several of the fundamental constraints~\cite{Bevan:2013kaa}.

Given the poor consistency between the inclusive
and exclusive determinations, any new determination of $|V_{ub}|$ is
then of the utmost importance. In this letter, we study the
semileptonic decay $\bar B_s\to K^+ \ell^-\bar\nu_\ell$. This decay channel is
expected to be observed at LHCb and Belle and it could be used to get an
independent determination of $|V_{ub}|$.

The semileptonic $\bar B_s\to K$ decay was analyzed in
Refs.~\cite{Li:2001yv,Duplancic:2008tk} using LCSR and the relevant
form factors were determined in the low $q^2$ region. Very recently,
preliminary lattice QCD (LQCD) estimates for those form factors for
higher $q^2$ values in the vicinity of $q^2_{\rm max}=(M_{B_s}-M_K)^2$
have become available~\cite{Bouchard:2013zda}, and a new wave of
theoretical studies on some QCD-motivated
models~\cite{Faustov:2013ima,Verma:2011yw, Su:2011eq, Wang:2012ab}
have appeared as well.  Parameterizations of the relevant form factor
$f^+$ are provided by the relativistic quark (RQM), covariant
light-front quark (LFQM) models and the perturbative (PQCD) approach
of Refs.~\cite{Faustov:2013ima}, \cite{Verma:2011yw} and
\cite{Wang:2012ab}, respectively. These were used in
Ref.~\cite{Meissner:2013pba} to compute and compare the different
predictions for the differential and partially integrated
decay widths for $\bar B_s\to K^+ \ell^-\bar\nu_\ell$, $\ell=e,\mu$ and
$\ell=\tau$ decays, as well as some forward-backward asymmetry and the
polarization fraction for the $\tau$ lepton. Conclusions however turned out to
be inconclusive, because of the large discrepancies among the
predictions from the different models considered.

Here we intend to obtain an improved description of the $f^+$ form factor
in the whole $q^2$ range accessible, $[0, q^2_{\rm max}] $, in
the decay and use it to predict the decay width ($\ell=e,\mu$) in
units of $|V_{ub}|^2$.  We  follow the strategy of earlier work in
Ref.~\cite{Albertus:2005ud}, where we analyzed the semileptonic
$B\to\pi$, $D\to\pi$ and $D_s\to K$ decays, and we  use the quark
model to evaluate the valence plus $\bar B^*$-pole contribution to the form
factors. In this way we will derive reliable $f^+$ form factor values
for high $q^2$, where the $\bar B^*$-pole contribution dominates, in good
agreement with the LQCD results reported in \cite{Bouchard:2013zda}
and the RQM calculation of  Ref.~\cite{Faustov:2013ima}.
Predictions in this region of the PQCD (LFQM) approach differ from ours by more
than a factor of 3 (10). Our estimates for $f^+$ are however not
accurate in the low $q^2$ region where the recoil of the final kaon is
large. On the other hand, LCSR results, though trustful in the
vicinity of $q^2=0$, cannot be used to describe $f^+$ above 10 GeV$^2$. 
We will then adopt the scheme of Refs.~\cite{Flynn:2006vr,
  Flynn:2007qd,Flynn:2007ii} and we shall take a
multiply subtracted Omn\`es functional ansatz for the dominant $f^+$
form factor and will make a combined fit to our quark model results in
the high $q^2$ region and to the LCSR results in the low $q^2$
region. The Omn\`es representation is employed to provide
a parameterization of the form factor constrained by unitarity and
analyticity properties.  In this way, we obtain a determination of the
form factor in accordance with LCSR and lattice results, that can be
used to determine the decay width. For low and intermediate $q^2$
values, discrepancies with the PQCD and LFQM approaches are not as
dramatic as in the vicinity of $q^2_{\rm max}$, though they are still
significant, while some disagreement with the RQM predictions show up
now,  that lead to a totally integrated width around 20\% larger in this
work than that reported in \cite{Faustov:2013ima}.

\section{Semileptonic $\bar B_s\to K$ decay}
For a $0^-\to 0^-$ transition, the weak hadronic matrix element can be parameterized as
\bea
\langle K^+,\,\vec
p_K\,|\bar\Psi_u(0)\gamma^\mu(1-\gamma_5)\Psi_b(0)|\bar B_s,\,\vec
p\,\rangle=\Big(P^\mu-q^\mu\frac{M_{B_s}^2-M_K^2}{q^2}\Big) f^+(q^2)
+q^\mu\frac{M_{B_s}^2-M_K^2}{q^2} f^0(q^2)
\eea
with $P=p+p_K,\,q=p-p_K$ and where $f^+(q^2), f^0(q^2)$ are form factors. In the case of
small lepton masses ($l=e,\mu$) the part proportional to $q^\mu$ gives a very
small contribution, when contracted with the leptonic current, and it can 
safely be neglected. Only the $f^+(q^2)$ form factor would
play a role in the decay. 

For zero lepton masses, the differential decay width is in fact given by
\bea
\frac{d\Gamma}{dq^2}=\frac{G_F^2}{192\pi^3}\,|V_{ub}|^2\,
\frac{\lambda^{3/2}(q^2,M_{B_s}^2,M_K^2)}{M_{B_s}^3}\,|f^+(q^2)|^2
\label{eq:dgdq2}
\eea
with $G_F=1.166378\times 10^{-5}\,$GeV$^{-2}$  the Fermi decay constant and
$|V_{ub}|$ the modulus of the corresponding Cabibbo-Kobayashi-Maskawa matrix
element. $\lambda$ is the K\"allen function defined as 
$\lambda(a,b,c)=a² +b^2+c^2-2ab-2ac-2bc$.

\subsection{Valence contribution to the form factors
}For a $\bar B_s$ meson initially at rest and taking $\vec q$ in the positive $Z$
direction ($\vec q=|\vec q\,|\vec k$), the valence quark model contribution to the form factors is
 evaluated as~\cite{Hernandez:2006gt}
\bea
f^+(q^2)&=&\frac{1}{2M_{B_s}}\,\left[
V^0(|\vec{q}\,|)+\frac{V^3(|\vec{q}\,|)}{|\vec{q}\,|}\,\left(E_K
(-\vec{q}\,)-M_{B_s}\right)\right]\nonumber\\
f^0(q^2)&=&\frac{1}{2M_{B_s}}\,\left\{
V^0(|\vec{q}\,|)\,\frac{q^2+M_{B_s}^2-M_K^2}{M_{B_s}^2-M_K^2}+
\frac{V^3(|\vec{q}\,|)}{|\vec{q}\,|}\,\left[
E_{K}
(-\vec{q}\,)\frac{q^2+M_{B_s}^2-M_K^2}{M_{B_s}^2-M_K^2}+M_{B_s}
\frac{q^2-M_{B_s}^2+M_K^2}{M_{B_s}^2-M_K^2}
\right]\right\}
\eea
with $V^0$ and $V^3$ the following vector matrix elements
\bea
V^0(|\vec{q}\,|)&=&\sqrt{2M_{B_s}2E_K(-\vec{q}\,)}\ \int\,d^3p\ \frac{1}{4\pi}
{\Phi}_K^*(|\vec{p}\,|)\,
{\Phi}_{B_s}\left(\bigg|\,\vec{p}-\frac{m_{s}}{m_{u}+m_{s}}
|\vec{q}\,|
\vec{k} \bigg|\right)\nonumber\\
&&\hspace{3cm}\sqrt{\frac{\widehat{E}_{u}\widehat{E}_{b}}{4E_{u}
E_{b}}}
\left(
1+\frac{(-\frac{m_{u}}{m_{u}+m_{s}}\,|\vec{q}\,|\vec{k}-\vec{p}\,)
\cdot(\frac{m_{s}}{m_{u}+m_{s}}\,|\vec{q}\,|\vec{k}-\vec{p}\,)}
{\widehat{E}_{u}\widehat{E}_{b}}
\right) \nonumber\\
V^3(|\vec{q}\,|)&=&\sqrt{2M_{B_s}2E_K(-\vec{q}\,)}\ \int\,d^3p\ \frac{1}{4\pi}
{\Phi}_{B_s}\left(\bigg|\,\vec{p}-\frac{m_{s}}{m_{u}+m_{s}}
|\vec{q}\,|
\vec{k} \bigg|\right)
 \nonumber\\
&&\hspace{3cm}\sqrt{\frac{\widehat{E}_{u}\widehat{E}_{b}}{4E_{u}
E_{b}}}
\left(\frac{\frac{m_{s}}{m_{u}+m_{s}}\,|\vec{q}\,|-p_z}{\widehat{E}_{b}}+
\frac{-\frac{m_{u}}{m_{u}+m_{s}}\,|\vec{q}\,|-p_z}{\widehat{E}_{u}}
\right) 
\eea
Here  $E_u=E_u(-\frac{m_{u}}{m_{u}+m_{s}}\,|\vec{q}\,|\vec k-\vec p\,),E_b=
E_b(\frac{m_{s}}{m_{u}+m_{s}}\,|\vec{q}\,|\vec k-\vec p\,)$ while 
$\widehat E_q=E_q+m_q$. The wave functions 
(Fourier transforms of the radial coordinate space $\bar B_s$ and $K$ meson wave
functions, which describe the relative dynamics of the
quark-antiquark pair) are evaluated using the AL1
interquark potential of Refs.~\cite{Semay:1994ht,
  SilvestreBrac:1996bg}. 
This potential contains
a linear confinement term plus $1/r$ and hyperfine terms coming from
one-gluon exchange. The masses and the rest of the parameters were fitted 
in Ref.~\cite{Semay:1994ht} to reproduce
the light and heavy-light meson spectra.

\begin{figure}[t]
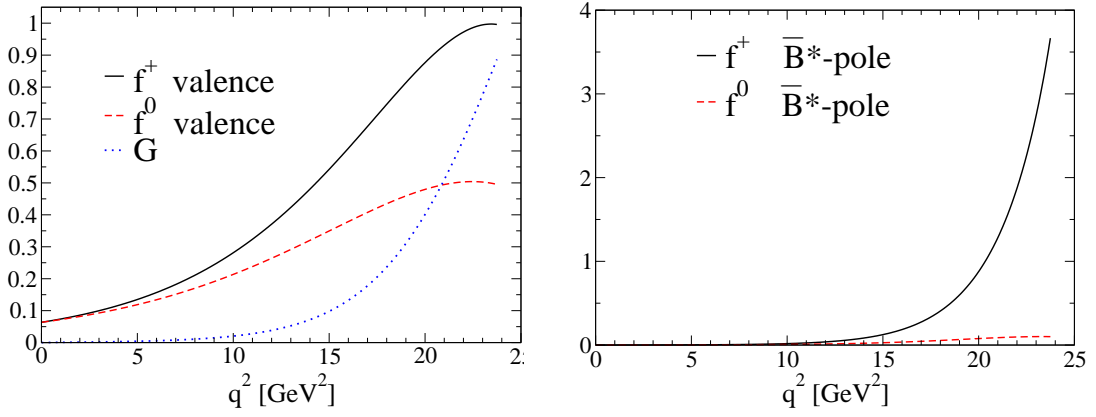

\centering
\resizebox{7cm}{!}{\includegraphics{nopolo.eps}}\hspace{.5cm}
\resizebox{6.75cm}{!}{\includegraphics{polo.eps}}
\caption{Left panel: Valence quark contribution to the $f^+(q^2)$ and 
$f^0(q^2)$ form factors and the $G(q^2)$ function as defined in Eq.(\ref{eq:g}).
 Right panel: $\bar B^*$-pole contribution to the $f^+(q^2)$ and 
$f^0(q^2)$ form factors evaluated in the constituent quark model.} 
\label{fig:ffnopolo}
\end{figure}

In the left panel of Fig~\ref{fig:ffnopolo} we show the results for the form factors 
thus obtained.
These form factors are not correct in the high $q^2$ region where the pole of
the $\bar B^*$ makes the largest contribution, and they are also incorrect at low $q^2$
where the recoil of the final meson is largest. We shall improve  their
behavior in both regions

\subsection{$\bar B^*$-pole contribution to the form factors.}
\begin{figure}[t]
\centering
\resizebox{!}{2.cm}{\includegraphics{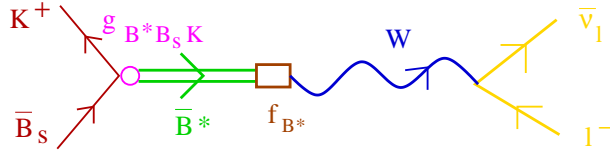}}
\caption{Feynman diagram corresponding to the $\bar B^*$-pole contribution to the
$\bar B_s\to K$ semileptonic decay.}
\label{fig:polo}
\end{figure}

The Feynman diagram for the $\bar B^*$-pole contribution to the decay process 
appears in Fig.~\ref{fig:polo}.
The corresponding weak matrix element is given by~\cite{Albertus:2005ud}
\bea
g_{B^*B_sK}(q^2)\sqrt{q^2}f_{B^*}(q^2)\,\frac{p_{K}^\mu-q^\mu(p_K\cdot q)/M_{B^*}^2}
{ M_{B^*}^2-q^2}
\eea
from where
\bea
f^{+}_{\rm pole}(q^2)&=&\frac{g_{B^*B_sK}(q^2)}2\sqrt{q^2}f_{B^*}(q^2)
\frac1{ M_{B^*}^2-q^2},\\
f^{0}_{\rm pole}(q^2)&=&\frac{g_{B^*B_sK}(q^2)}2\sqrt{q^2}f_{B^*}(q^2)
\frac{M_{B_s}^2-M_K^2-q^2}{(M_{B_s}^2-M_K^2) M_{B^*}^2}
\eea
Within the quark model we have that~\cite{Albertus:2005vd}
\bea
f_{B^*}(q^2)&=&\frac{\sqrt6}{(q^2)^{1/4}\,\pi}\int_0^\infty d|\vec p\,|
\Phi_{B_u^*}(|\vec p\,|)\,|\vec p\,|^2
\sqrt{\frac{\widehat E_b\widehat E_u}{4E_b
E_u}}\,\Big(1+\frac{|\vec p\,|^2}{3\widehat E_b
\widehat E_u}\Big)\\
&=&f_{B^*}\sqrt{\frac{ M_{B^*}}{\sqrt{q^2}}}
\eea
with $E_q=E_q(|\vec p\,|)=\sqrt{m_q^2+\vec{p}^{\,2}},\, \widehat E_q=E_q+m_q$, and where
$f_{B^*}$ is the 
on-shell decay constant for which we get $f_{B^*}=151\,$MeV~\cite{Albertus:2005vd}.
Finally $g_{B^*B_sK}(q^2)$ is obtained as explained in  Ref.~\cite{Albertus:2005vd}.
We write it  as
\bea
g_{B^*B_sK}(q^2)=g_{B^*B_sK}\,G(q^2)
\label{eq:g}
\eea where $g_{B^*B_sK}$ is the $B^*B_sK$ coupling constant evaluated
at $q^2=M^2_{B^*}$ but in the chiral limit (zero kaon mass). We get
for its value $g_{B^*B_sK}=49.88$. The $G(q^2)$ function is a
dimensionless hadronic factor normalized to one at $q^2= M_{B^*}^2$,
which accounts for the $q^2$ dependence of $\bar B_s\to \bar B^* K$
amplitude. It is shown in the left panel of Fig.~\ref{fig:ffnopolo}.
The product $g_{B^*B_sK}\,f_{B^*}$ is given in our model as 
\bea
g_{B^*B_sK}\,f_{B^*}= 7.53\,{\rm GeV} 
\eea 
This value is too large
compared to the LCSR determination of Ref.~\cite{Li:2001yv}. There the
authors get values in the range
$g_{B^*B_sK}\,f_{B^*}=3.57-4.19\,$GeV. On the other hand one can make
use of SU(3) symmetry to get the relation~\cite{Colangelo:2002dg}
 \bea
g_{B^*B_s K}\,f_{B^*}=g_{B^*B\pi}\,f_{B^*}
\sqrt{\frac{M_{B_s}}{M_B}}\frac{f_\pi}{f_K}
 \eea 
Lattice data for
$f_{B^*}$~\cite{Bowler:2000xw} and $g_{B^*B\pi}$~\cite{Abada:2003un}
gives $g_{B^*B\pi}\,f_{B^*}=8.9\pm 2.2\,$GeV, from where SU(3)
symmetry will predict \bea g_{B^*B_s K}\,f_{B^*}=7.49\pm1.85\,{\rm
  GeV} \eea in good agreement with our determination. We shall use
this latter result and the error will serve to evaluate the
uncertainties in our quark model calculation. The $\bar B^*$-pole
contribution to the form factors using
$g_{B^*B_sK}\,f_{B^*}=7.49\,{\rm GeV}$ is presented in the right panel
of Fig.~\ref{fig:ffnopolo}.

\begin{figure}[t]
\centering
\resizebox{!}{5.cm}{\includegraphics{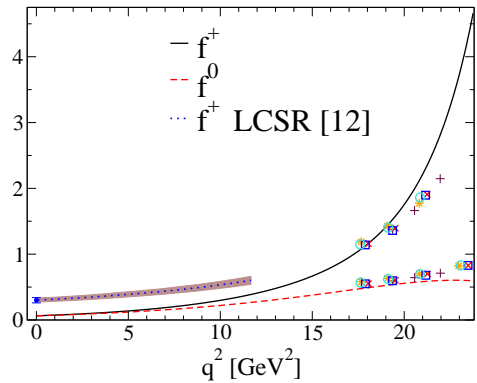}}
\caption{Total, valence plus $\bar B^*$-pole
contributions, $f^+(q^2)$ and $f^0(q^2)$ form factors. We also show the 
 results obtained in the LCSR calculation
of Ref.~\cite{Duplancic:2008tk} (dotted-line plus error band) 
and different preliminary lattice data  in the high $q^2$ region
reported in Ref.~\cite{Bouchard:2013zda}.}
\label{fig:total}
\end{figure}

The total quark model form factors, valence plus $\bar B^*$-pole
contributions, are shown in Fig.~\ref{fig:total}. There, we also show
the results of the LCSR calculation of Ref.~\cite{Duplancic:2008tk} (very similar
results, not shown, are obtained in Ref.~\cite{Li:2001yv}), and
preliminary LQCD results for high $q^2$ obtained in Ref.~\cite{Bouchard:2013zda}
with different lattice configurations. In the LCSR
approach of Ref.~\cite{Duplancic:2008tk} a value of
$f^+(0)=0.30^{+0.04}_{-0.03}$ is reported and we have assumed a similar 10\% error
on the LCSR form factor at larger $q^2$  that we show as
an error band in Fig.~\ref{fig:total}.
Our results are compatible with lattice data  if one 
takes into account the
uncertainties in the $g_{B^*B_sK}\,f_{B^*}$ value. However,  the disagreement
with the LCSR results in the low $q^2$ region can not be corrected in
that way. The quark model evaluation is not appropriate for low $q^2$ where
the kaon recoil is large. In the next subsection we shall use the Omn\`es
representation to combine our quark model results,
that we consider to be reliable in the high $q^2$ region, with the LCSR 
results in the low $q^2$ part. In this way we
shall get a realistic description of the form factor for all $q^2$ values
allowed in the decay.
\subsection{Omn\`es representation of the $f^+$ form factor}
Here we shall use the multiply subtracted Omn\`es representation of the 
$f^+$ form factor as 
discussed in Ref.~\cite{Flynn:2007qd}, where the $\bar B^*$-pole has been  made explicit
\begin{eqnarray}
f^+(q^2)&\approx&\frac{1}{M^2_{B^*}-q^2}\prod_{j=0}^n\Big[f^+(q_j^2) 
\Big(M^2_{B^*}-q^2_j\Big)\Big]^{\alpha_j(q^2)}
\label{eq:omnes}\\
\alpha_j(q^2) &=& \prod_{j\ne k=0}^n \frac{q^2-q_K^2}{q^2_j-q^2_k}
\end{eqnarray}
with $q^2 < s_{\rm th}=(M_{B_s}+M_K)^2$ and $q_0, \cdots q_n^2 \in
]-\infty, s_{\rm th}[$, the $q^2-$values where the $(n+1)$ subtractions are
    considered. In addition, $f^+(q_j^2)$ are the values that the form
    factor takes at the subtraction points. The Omn\`es representation
    above emerges from Watson's theorem that states, 
\bea
\frac{f^+(s+i\epsilon)}{f^+(s-i\epsilon)} = e^{2i\delta(s)}, \qquad s \ge
s_{\rm th}
\eea
where $\delta(s)$ is the phase-shift for elastic $K \bar B_s \to K
\bar B_s$ scattering in the total angular momentum $J=1$ channel.  In
principle, the Omn\`es representation requires as an input the phase
shift plus the form factor at $(n+1)$ positions $\{q^2_i\}$ values
below the $K\bar B_s$ threshold. For sufficiently many subtractions,
the phase shift $\delta (s)$ can be approximated by its value at
threshold\footnote{It is set to zero, with the help of Levinson's
  theorem (see the detailed discussion in Ref.~\cite{Flynn:2007qd}).}
leading to the approximate representation of
Eq.~(\ref{eq:omnes}). This amounts to finding an interpolating
polynomial for $\ln\left[ (M^2_{B^*}-q^2)f^+(q^2) \right]$ passing
through the points $\ln\left[ (M^2_{B^*}-q^2_i)f^+(q^2_i) \right]$.
While one could always propose a parametrization using an
interpolating polynomial for $\ln\left[g(q^2)f^+(q^2)\right]$ for a
suitable function $g(q^2)$, the derivation in ~\cite{Flynn:2007qd}
using the Omn\`es representation shows that taking $g(q^2) =
(M^2_{B^*}-q^2)$ is physically motivated.

The approach that we shall follow is the one in
Ref.~\cite{Flynn:2007ii}. Taking for $q^2_j$ the four different values
$0,\,q^2_{\rm max}/3,\,2q^2_{\rm max}/3$ and $q^2_{\rm max}$, we treat
$f^+(q_j^2)$ as free parameters and make a combined $\chi^2-$fit to
our quark model results and the LCSR predictions of
Ref.~\cite{Duplancic:2008tk} in the high and low $q^2$ regions,
respectively.  We take LCSR values for $f^+$ at $q^2=0,2,4,6,8$
and 10 GeV$^2$ with a 10\% relative error, as mentioned
above, while we fit to our quark model results for
$q^2=19,20,21,22$ and 23 GeV$^2$, and assign to these points the error that derives
from using $g_{B^*B_sK}\,f_{B^*}=7.49\pm1.85\,{\rm GeV}$.

\begin{figure}[t]
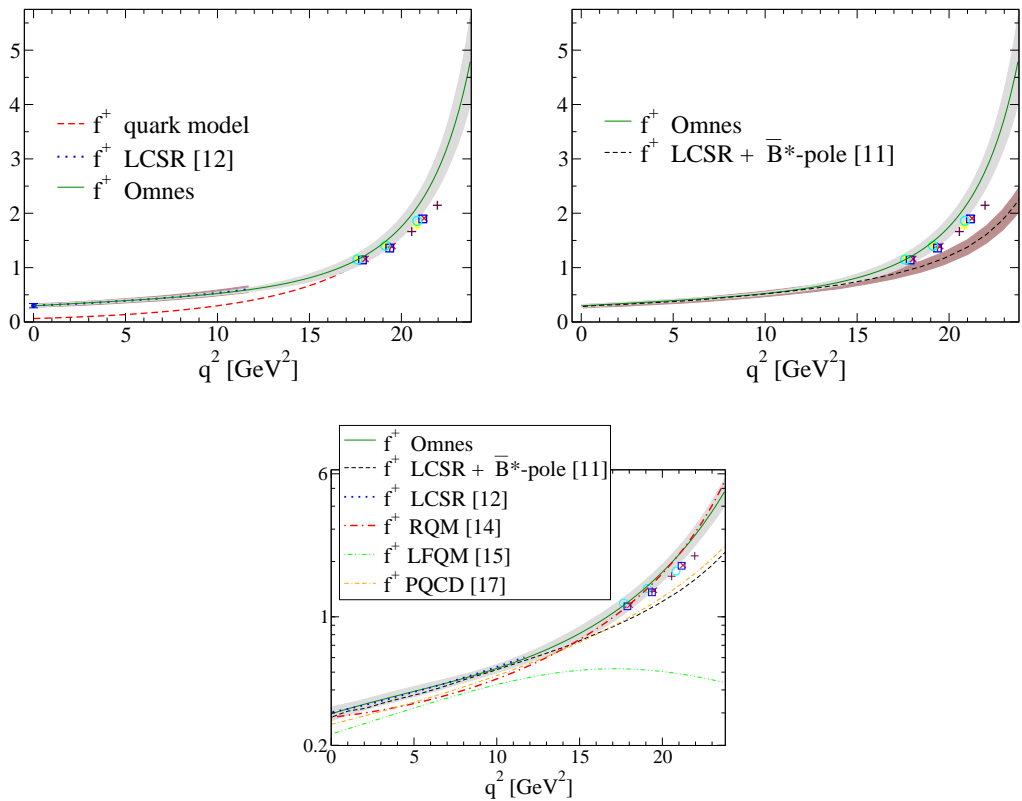

\centering
\resizebox{!}{5.cm}{\includegraphics{omnes2_paper.eps}}\hspace{1cm}
\resizebox{!}{5.cm}{\includegraphics{omnes3_paper.eps}}\vspace{0.5cm}
\resizebox{!}{5.cm}{\includegraphics{omnes4_paper.eps}}
\caption{Upper-left panel: $f^+(q^2)$ evaluated in the quark model
  (red dashed line) and
  improved at lower $q^2$ by means of the Omn\`es representation
  (solid line plus 68\% confidence level band).  We also show 
   the LCSR results of 
  Ref.~\cite{Duplancic:2008tk} (blue dotted-line+10\% error band) and different preliminary
  lattice data taken from Ref.~\cite{Bouchard:2013zda}. Upper-right panel: We 
  compare our final result (solid line plus 68\% confidence level band) 
  with the fit in Ref.~\cite{Li:2001yv} 
  (dashed-line). For the
latter a 10\% error band is also displayed. Lattice data points taken from
Ref.~\cite{Bouchard:2013zda} are also shown. Lower panel: Global comparison of
our final result for the $f^+$ form factor with different calculations using
LCSR~\cite{Duplancic:2008tk}, LCSR$+\bar B^*$-pole fit~\cite{Li:2001yv}, 
RQM~\cite{Faustov:2013ima}, LFQM~\cite{Verma:2011yw} and
PQCD~\cite{Wang:2012ab}. Lattice data from Ref.~\cite{Bouchard:2013zda} are also
shown for comparison.}
\label{fig:omnes2}
\end{figure}

The outcome of the fit is
\bea
f^+(0)&=&0.297\pm0.026,\nonumber\\
f^+(q^2_{\rm max}/3)&=&0.460\pm0.025,\nonumber\\
f^+(2\,q^2_{\rm max}/3)&=&0.896\pm 0.084,\nonumber\\
f^+(q^2_{\rm max})&=&4.792\pm0.808
\label{eq:fit} 
\eea
and the corresponding form factor, together with a 68\% confidence level band,
 is depicted in the upper-left panel of Fig.~\ref{fig:omnes2}. The procedure to build the
68\% confidence level band for the form factor is the following: We generate
 a 1000 sets of 
 $(f^+(0),f^+(q^2_{\rm max}/3),f^+(2\,q^2_{\rm max}/3),f^+(q^2_{\rm max}))$
 values
 assuming
 an uncorrelated four-dimensional Gaussian distribution for which the 
 central values and the standard deviations are taken from Eq.~(\ref{eq:fit}).
 In this way we generate a 1000 different $f^+$ form factors. The 68\%
 confidence level band is built discarding for each $q^2$  the 16\% largest
 and 16\% lowest values of the form factor. 

A different fit was performed in Ref.~\cite{Li:2001yv}. There, a value
of $g_{B^*B_sK}f_{B^*}=3.88\pm0.31\,{\rm GeV}$ was used and as a
consequence the form factor extracted in \cite{Li:2001yv} is very
different from ours for large $q^2-$values, where the $\bar B^*$-pole
contribution dominates. A comparison of the two form factors is given
in the upper-right panel of Fig.~\ref{fig:omnes2}.

In the lower panel of Fig.~\ref{fig:omnes2} we compare the $f^+$ form factor
as calculated in different approaches. The LCSR calculation of 
Ref.~\cite{Duplancic:2008tk} only provides results up to $q^2=10$\,GeV$^2$ and
the PQCD
calculation of Ref.~\cite{Wang:2012ab} does not include a $\bar B^*$-pole 
contribution and then its form factor
is not reliable in the high $q^2$ region. All other calculations do include the
$\bar B^*$-pole mechanism, but they differ in its strength. The RQM
calculation of  Ref.~\cite{Faustov:2013ima} provides a  result similar to ours in
the region of $q^2$ dominated by the $\bar B^*$-pole,  
 while  Refs.~\cite{Li:2001yv,Wang:2012ab} give smaller values. Lattice data
  in the high $q^2$ region will be most
valuable to decide which of the two sets of calculations is more correct. 
As seen in the figure there are also differences in the low $q^2$ part that 
will get amplified in the differential decay width.
\subsection{Prediction for the decay width}
\begin{figure}[t]
\centering
\resizebox{!}{5.cm}{\includegraphics{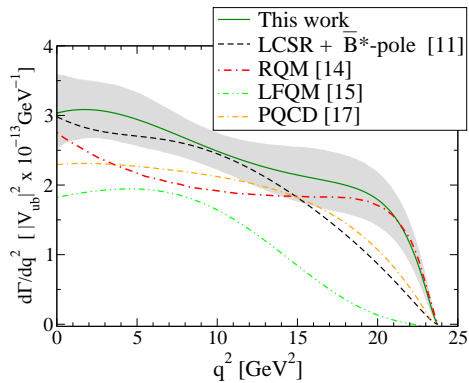}}
\caption{Differential decay width obtained in this work with the Omn\`es fit (solid line
plus 68\% confidence level band) and in LCSR$+\bar B^*$-pole fit~\cite{Li:2001yv}, 
RQM~\cite{Faustov:2013ima}, LFQM~\cite{Verma:2011yw} and
PQCD~\cite{Wang:2012ab} approaches.}
\label{fig:omnes3}
\end{figure}%
With the Omn\`es form factor,  we evaluate the differential decay width
that is displayed in  Fig.~\ref{fig:omnes3}, together with
 its 68\% confidence level band.  We
also show the differential decay width that derives from the calculations in 
Refs.~\cite{Li:2001yv,Faustov:2013ima,Verma:2011yw,Wang:2012ab}. The differences
present in the form factors at low and high $q^2$ have a clear reflection 
here.

For the integrated decay width we finally obtain 
\bea
\Gamma(\bar B_s\to K^+\ell^-\bar\nu_\ell)=(5.45^{+0.83}_{-0.80})|V_{ub}|^2\times 10^{-9}\,{\rm
MeV} \label{eq:final}
\eea
A comparison with the other approaches is given in
Table~\ref{tab:dw}. The calculations in
Refs.~\cite{Li:2001yv,Faustov:2013ima} predict very similar results,
even though their form factors deviate both in the low and high $q^2$
region, but whose effects compensate in the integrated width. The
result of the PQCD calculation of Ref.~\cite{Wang:2012ab} is also
similar, but in that case the uncertainty, as quoted in
Ref.~\cite{Meissner:2013pba}, is around 50\%. A smaller result is
given in the LFQM calculation of Ref.~\cite{Verma:2011yw}.  This is in
part a reflection of the fact that no $\bar B^*$- pole contribution is
included in that approach. Our result is the largest although we agree with
Refs.~\cite{Li:2001yv,Faustov:2013ima,Wang:2012ab} within
uncertainties.
\begin{table}
\begin{tabular}{l|ccccc}\hline\hline
&This work&LCSR+$\bar B^*$-pole&RQM&LFQM&PQCD\\
&&\cite{Li:2001yv}&\cite{Faustov:2013ima}&\cite{Verma:2011yw}&\cite{Wang:2012ab}\\
\hline
$\Gamma\ [\,|V_{ub}|^2\times 10^{-9}\,{\rm MeV}]$&$5.45^{+0.83}_{-0.80}$
&$4.63^{+0.97}_{-0.88}$&$4.50\pm0.55$&
$3.17\pm0.24$&$4.2\pm2.1$\\\hline\hline
\end{tabular}
\caption{Decay width in units of $|V_{ub}|^2\times 10^{-9}\,{\rm MeV}$
  from several approaches. For the
result of Ref.~\cite{Li:2001yv} we have propagated a 10\% uncertainty in the
form factor. Results for Refs.~\cite{Faustov:2013ima,Verma:2011yw,Wang:2012ab}
have been adapted from Table IV in Ref.~\cite{Meissner:2013pba}.}
\label{tab:dw}
\end{table}

One might be tempted to combine\footnote{We symmetrize the errors displayed 
in Table
  \ref{tab:dw} and use the maximum likelihood method
for which $\Gamma_{\rm
    avg}/\sigma^2 = \sum_i \Gamma_i/\sigma_i^2 $, with $1/\sigma^2=
  \sum_i 1/\sigma_i^2$. } our result with the ones in
Refs.~\cite{Li:2001yv,Faustov:2013ima}. In that case, one would get
\bea \Gamma(\bar B_s\to
K^+\ell^+\bar\nu_\ell)=(4.77\pm0.41)|V_{ub}|^2\times 10^{-9}\,{\rm
  MeV} \eea
Once experimental data is available, the above result may
be used to obtain the value of $|V_{ub}|$ with a theoretical error of
the order of 5\%. Nevertheless, we believe that our estimate for
$\Gamma(\bar B_s\to K^+\ell^-\bar\nu_\ell)$ in Eq.~(\ref{eq:final}) is
more accurate.  The RQM approach at low and intermediate values of
$q^2$ should not be as appropriated as the LCSR scheme, whose input is
included in our combined scheme. Conversely, our quark model
predictions for $f^+$ agree remarkably well with those of the RQM of
Ref.~\cite{Faustov:2013ima} at high $q^2$ values, and are
significantly larger than those provided by the $\bar B^*$ pole
contribution assumed in \cite{Li:2001yv}. Indeed, the value for
$g_{B^*B_sK}f_{B^*}$ used in that work strongly disagrees, both with
our predictions and with the existing LQCD data. Moreover, the
agreement showed in Table \ref{tab:dw} among
Refs.~\cite{Li:2001yv,Faustov:2013ima} is just a coincidence, because
their respective predictions for $f^+(q^2)$ and $d\Gamma/dq^2$ in
Figs. \ref{fig:omnes2} (bottom panel) and \ref{fig:omnes3}
significantly differ for most of the available phase space.

\section{Summary}

We have studied the form factor $f^+$ for the semileptonic $\bar
B_s\to K^+\ell^-\bar\nu_\ell$ decay within an Omn\`es scheme, which
incorporates unitarity and analyticity constrains and it makes
possible to combine quark model and LCSR results in the high and low
$q^2$ regions, respectively. We predict $\Gamma(\bar B_s\to
K^+\ell^-\bar\nu_\ell)$ with a theoretical uncertainty of the order of
15\%, inherited from the 10\% and 25\% errors on the LCSR and quark
model inputs for $f^+$, respectively. Uncertainties on the predicted
differential width (Fig.~\ref{fig:omnes3}) turn also to be moderately
small.  Once experimental data is available, our prediction for
$\Gamma(\bar B_s\to K^+\ell^-\bar\nu_\ell)$ in Eq.~(\ref{eq:final}) or
 $d\Gamma/dq^2$  could be used to
determine $|V_{ub}|$ with a theoretical error of the order of 7\%,
which will be reduced by improved LCSR and LQCD results. In
particular, LQCD simulations could not only provide the form factor
$f^+$, that could be directly fitted, but also a more accurate
determination of $g_{B^*B_sK}f_{B^*}$, which is the major source of
uncertainty in the quark model input included in the combined Omn\`es
analysis. Experimental data for this reaction is expected from the LHCb and
Belle Collaborations in the near future.

\begin{acknowledgments}
 This research was supported by  the Spanish Ministerio de Econom\'{\i}a y 
 Competitividad and European FEDER funds
under Contracts Nos. FPA2010-21750-C02-02,  FIS2011-28853-C02-02,  
and the Spanish Consolider-Ingenio 2010 Programme CPAN (CSD2007-00042), by Generalitat
Valenciana under Contract No. PROMETEO/20090090, by Junta de Andalucia under
Contract No. FQM-225,
 by the EU HadronPhysics3 project, Grant Agreement
No. 283286,  and by the
University of Granada start-up Project for Young Researches contract No. PYR-2014-1.
C.A. wishes to acknowledge a CPAN postdoctoral contract and C.H.-D. thanks the support 
of the JAE-CSIC Program.
  \end{acknowledgments}

\end{document}